# Mechanical properties of electrodeposited amorphous/crystalline multilayer structures in the Fe-P system

Timo Müller[a*], Andrea Bachmaier[a], Ruth Konetschnik[b], Thomas Schöberl[a], Reinhard Pippan[a]

[a]Erich Schmid Institute of Materials Science, Austrian Academy of Sciences, Leoben, Austria
[b]Department Materials Physics, Montanuniversität Leoben, Austria

**Abstract**
Amorphous/crystalline multilayer structures of Fe-P alloys were deposited electrochemically using the single bath technique. Hall-Petch behavior of microhardness with respect to sublayer thickness was observed down to a sublayer thickness of 15 nm. For thinner sublayers, a hardness plateau was obtained. The transition at a sublayer thickness of 15 nm coincides with the loss of the multilayer structure as observed in transmission electron microscopy. The transition is a possible result of a change in the amorphous to crystalline sublayer thickness ratio and the interface roughness development during the deposition process. Additionally, crack deflection at the interfaces was observed for the layered structures with small sublayer thickness in microbending experiments.

**Keywords:** electrodeposition; multilayer; hardness; Hall-Petch behavior; iron alloys

1. Introduction

Lamellar structures with lamella widths in the submicron to nanometer regime have attracted much attention due to their unique properties, for example concerning magnetism [1-3] or mechanical strength [4]. The mechanical strength is generally observed to increase with decreasing lamella thickness for crystalline structures. This is equivalent to the Hall-Petch behavior in bulk materials [5, 6]. Indeed, the size dependency for lamellar structures is well described in many cases using the Hall-Petch equation with the sublayer thickness instead of the grain size [7-10]. In other studies, a linear behavior of hardness or yield stress as a function of the reciprocal sublayer thickness was observed [11, 12]. Independent of the exact relationship of strength and sublayer thickness, no further increase or even a decrease of strength is observed below a certain sublayer thickness, which is typically between 2 and 50 nm [8-10, 13]. Various explanations for this phenomenon, which are mainly based on a change of deformation mechanisms on the nanometer scale, are discussed in literature [9, 10].

On the contrary, no clear dependence of the strength on the layer thickness is reported for amorphous layers [14]. This can be attributed to the different deformation mechanism of amorphous structures which is based on shear transformation zones as the carriers of plastic deformation [15]. Multilayer structures of alternating amorphous and crystalline sublayers are promising candidates for high strength materials since the combination of the different deformation mechanisms allows materials with high strength, strain hardening and reduced tendency for strain localization.

Most previous studies on the mechanical properties of amorphous/crystalline multilayer structures have been conducted on sputtered Cu/Cu-Zr multilayer structures [16-18]. It was found that the presence of the crystalline layers prevents the formation of catastrophic shear bands in the amorphous phase which was attributed to a minimum size that is necessary to induce shear localization [16]. Similar results were obtained by Donohue et al. for sputtered Cu/Pd-Si multilayer structures [19].

Multilayer structures can also be produced using electrochemical deposition. Two different techniques for electrodeposition of such structures exist [7]. In the dual bath technique, different electrolytes are used for each set of sublayers and the sample is transferred between the baths after deposition of each sublayer. This technique allows a large variety of multilayer systems, but is technologically challenging due to possible contamination during the transfer between the baths [7, 20]. The other technique, which is used in the present study, is the single bath technique. The complete multilayer structures are deposited from one electrolyte by changing one deposition parameter periodically during the deposition process. Thus, the technique is restricted to systems, in which all sublayer structures can be deposited from the same bath changing only one deposition parameter such as current density [7]. Both techniques have been applied to many systems, also including amorphous/crystalline multilayers [21-24].

Although later and less intensively investigated than the analogous systems Co-P and especially Ni-P [25], electrodeposition of Fe-P alloys is well known and various properties of such alloys have been examined. The structure of electrodeposited Fe-P alloys depends on their phosphorus content. For more than about 15 to 22 at.-% P an amorphous material is formed, whereas for smaller phosphorus contents the alloy crystallizes as interstitial solid solution of body-centered cubic structure [26-30]. Regularly, a mixture of crystalline and amorphous structures is reported for intermediate phosphorus contents [27, 28]. Similar results are also known for electroless Fe-P alloys [31, 32]. The phosphorus content of the deposits depends on many deposition parameters, such as bath composition, current density and temperature [27, 30, 33]. The composition range of amorphous structures can be extended towards lower phosphorus contents using pulsed deposition current [34, 35].

Although the dependency of the structure on the phosphorus content provides the possibility to produce layered structures with periodically changing phosphorus contents using the single bath technique, only a few studies exist on this topic, which are even restricted to the analogical system

Ni-P. Elias et al. deposited amorphous/crystalline multilayer samples in the Ni-P system and proved an improved corrosion resistance as compared to homogeneous coatings [36]. The Ni-P systems was also used as a model system during the improvement of the dual bath technique [20, 21] and diffusion measurements were performed on amorphous/crystalline multilayers produced with this technique [22]. Besides, indications of multilayer structures were found in Ni-P deposits prepared under constant potential or current density, respectively [37, 38]. For the Fe-P system, Kamei and Maehada investigated pulsed electrodeposition, but reported multilayer structures only for the Fe-Cu-P system [34]. Thus, no study on deposition of amorphous/crystalline multilayers of Fe-P alloys and no investigations on the mechanical properties of amorphous/crystalline electrodeposits in general exist to the authors' knowledge.

In this work, Fe-P multilayer structures are prepared using the single bath technique. The use of electrodeposition allows the production of samples with a significantly larger total thickness as compared to previous work on sputtered films [18]. The deposition parameters are chosen in such a way that both amorphous and crystalline sublayers can be obtained by changing the current density. The dependence of the mechanical properties of these structures on the sublayer thickness is evaluated using microhardness, nanoindentation, and microbending testing. Additionally, direct current samples are investigated to get more insights into the properties of the two phases combined in the multilayer structures.

2. Experimental details

2.1. Electrodeposition

The multilayer structures were deposited electrochemically on mechanically polished polycrystalline copper plates of 12 mm diameter. The bath composition was 160 g/l iron(II) sulfate heptahydrate, 12 g/l sodium chloride, 20 g/l ammonium chloride, 15 g/l sodium hypophosphite hydrate, 0.2 g/l sodium dodecyl sulfate. Deposition was carried out at 60°C and a pH of 2.35±0.05 was adjusted both before and during the experiments using sulfuric acid. An iron rod was used as soluble counter electrode. All deposition experiments were performed under galvanostatic control using an IPS Jaissle PGU OEM-2A-MI potentiostat. The current density was switched between -10 mA/cm² and -80 mA/cm² to obtain amorphous and crystalline layers, respectively. For all samples, except where noted otherwise, an identical nominal layer thickness for the amorphous and the crystalline layers was chosen. This nominal layer thickness was varied over a large range from 10 nm to 800 nm for different samples. The deposition times for each layer were chosen accordingly using Faraday's law. The total layer thickness was 50 μm for all samples.

2.2. Sample characterization

Cross-sections were investigated using scanning and transmission electron microscopy (SEM and TEM). SEM samples were polished with water-free suspensions. For investigations of the multilayer structures, they were subsequently etched for 10 seconds in a diluted nital solution consisting of 0.25 vol.-% nitric acid in ethanol. SEM investigations were carried out using a Zeiss LEO 1525 instrument, which was also used for energy-dispersive X-ray spectroscopy (EDX). A Philips CM12 machine was used for TEM analysis on cross-sections prepared by mechanical polishing and subsequent thinning in a focused ion beam machine. High-resolution TEM and EDX were performed on a JEOL 2100F machine.

X-ray diffraction (XRD) measurements were performed on the as-deposited film surface using a Rigaku SmartLab instrument with CuKα radiation. Vickers microhardness testing was also carried out on the as-deposited surface with a load of 100 g (HV 0.1). Nanoindentation was performed with a diamond cube corner tip on water-free polished surfaces both parallel and perpendicular to the substrate surface using a Hysitron Triboscope with a maximal load of 6 mN.

2.4. Micromechanical testing

Microbending beams were prepared as described in [39] using a HITACHI E-3500 ion polisher and a Zeiss LEO 1540XB focused ion beam machine. The cuboidal bending beams were approximately 30 μm long, 9 μm broad and 5 μm thick after the final polishing with a $Ga^+$ ion current of 200 pA. The loading axis was perpendicular to the substrate surface. The beams were bent until catastrophic failure under displacement control with a velocity of 1 μm per minute. The bending experiments were carried out using an ASMEC UNAT-SEM microindenter in a Zeiss LEO 982 SEM.

3. Results

3.1. Comparative samples without multilayer structure

To prove the suitability of the system for amorphous/crystalline multilayer deposition, direct current deposits were prepared with the current densities that were later used for the two sets of sublayers in the multilayer samples. The XRD patterns of these deposits reveal the mainly amorphous and crystalline structure obtained at -10 mA/cm² and -80 mA/cm², respectively (Fig. 1). However, the XRD pattern of the amorphous structure contains some small Bragg peaks. Most of them can be attributed to the copper substrate, but the peak at 2θ = 82° seems to originate from bcc Fe-P crystallites. Nevertheless, these deposition conditions are a suitable candidate for the deposition of amorphous sublayers in a multilayer structure since the prolonged deposition with low current density for the direct current samples makes the formation of crystals more probable due to the depletion of phosphorus in the bath. Table 1 shows the hardness and the phosphorus contents for the direct current deposits. The difference in hardness obtained from microhardness testing and nanoindentation is attributed to the different indenter geometries, i.e. a Vickers indenter and a cube corner tip, respectively, in combination with the well-known indentation size effect [40]. The phosphorus contents correspond well with the compositions of amorphous and crystalline Fe-P electrodeposits according to literature [26-30].

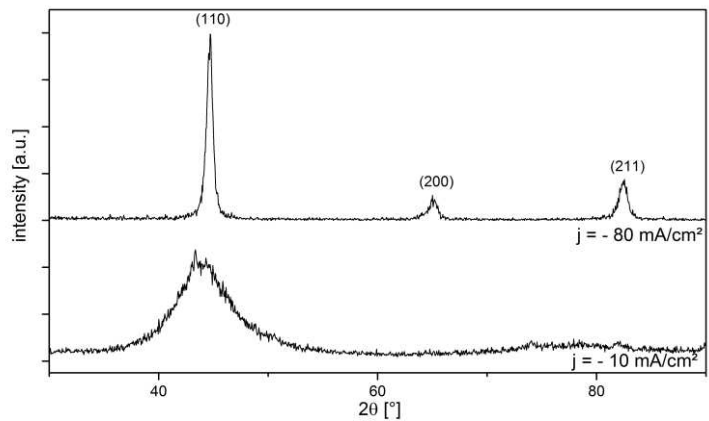

Fig. 1: XRD patterns of direct current deposits prepared with current densities of -80 mA/cm² and -10 mA/cm² resulting in a crystalline and amorphous structure, respectively.

| Current density [mA/cm²] | -10 | -80 |
|---|---|---|
| Structure | amorphous | crystalline |
| Microhardness [GPa] | 4.65 ± 0.18 | 4.94 ± 0.18 |
| Nanoindentation hardness [GPa] | 6.07 ± 0.04 | 8.70 ± 0.14 |
| at.-% P | 19.3 ± 0.1 | 6.8 ± 0.2 |

Table1: Hardness, phosphorus content and structure of direct current deposits. The phosphorus contents were determined using EDX in the SEM. Nanoindentation was performed on cross-sections.

3.2. Structural characterization of multilayer samples

The total thickness of the deposits was measured from SEM micrographs and the cathodic current efficiency was calculated using Faraday's law assuming that only the reduction of ferrous ions to iron atoms takes place at the cathode. The cathodic current efficiency scatters in a range from about 30 to 60 % without any obvious dependency on the sublayer thickness or any other measured deposition parameter. The average sublayer thickness for each sample was calculated as the product of the nominal sublayer thickness and the cathodic current efficiency obtained from the total film thickness measured in the SEM, assuming equivalent thickness of amorphous and crystalline sublayers.

Before etching, no sublayer structure was observed in the SEM. However, after etching the multilayer structure becomes visible over the complete range of sublayer thicknesses of this study (Fig. 2). Only for sublayer thicknesses of less than 15 nm the multilayer structure was usually not observed over the whole sample, but only in some regions. Besides, in thin multilayer structures also delamination, i.e. cracking parallel to the layer orientation, was observed (Fig. 2d). However, no cracks were observed on the same samples before etching.

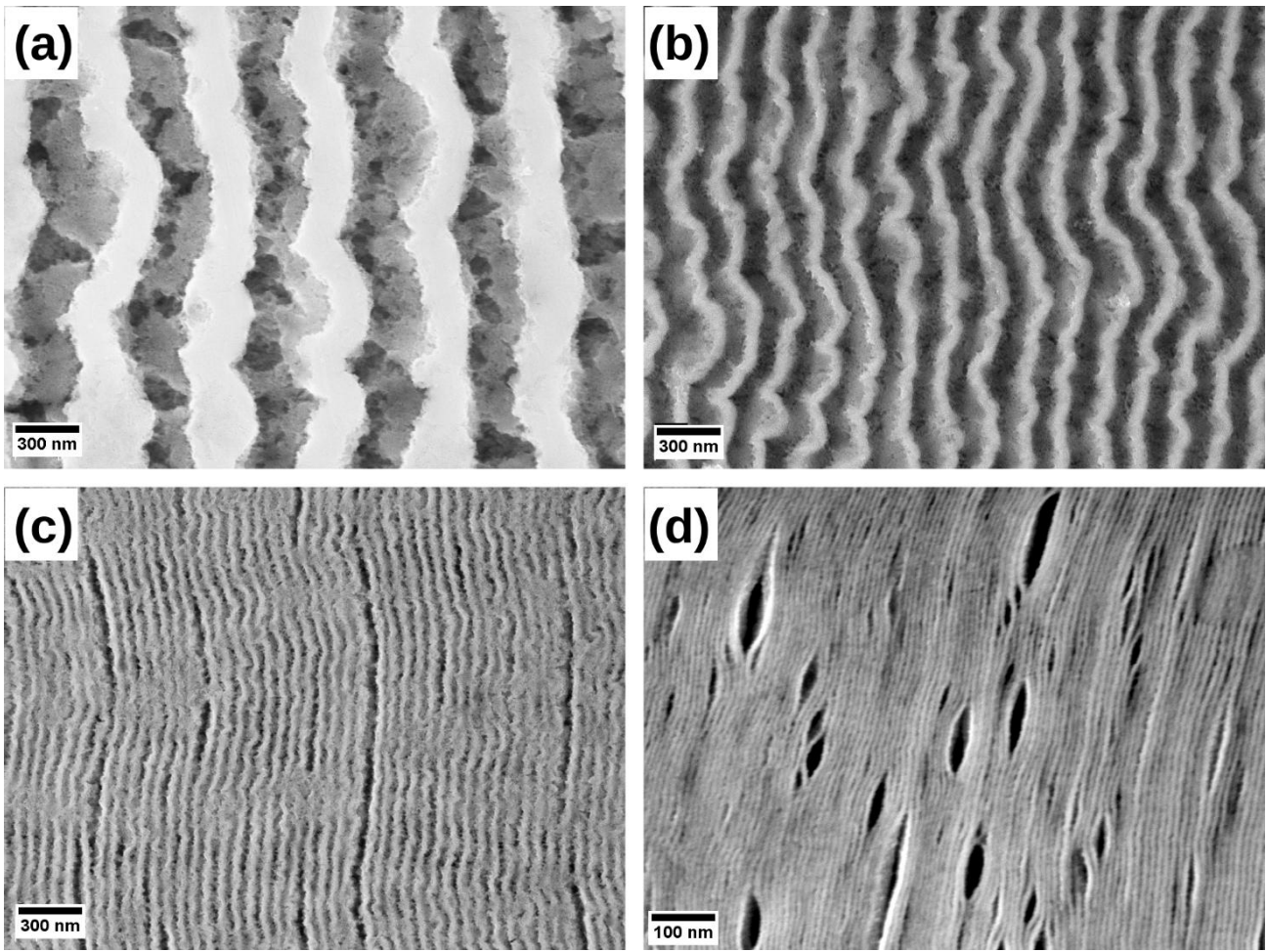

Fig. 2: SEM micrographs of etched multilayer samples with (a) 250 nm, (b) 100 nm, (c) 30 nm, and (d) 5 nm average sublayer thickness.

Amorphous and crystalline structures can be differentiated in the SEM since one set of sublayers is almost unaffected by etching, whereas the other one is attacked by the etching solution and shows a substructure indicating the presence of crystalline grains (Fig. 2a). The amorphous and crystalline structure of the sublayers was proven by TEM for selected samples (Fig. 3). For a sample with 250 nm average sublayer thickness, the area diffraction patterns recorded in the two sets of sublayers show their crystalline or amorphous structure, respectively (Fig. 3a). High-resolution TEM performed with samples of 290 nm (Fig. 3b) and 30 nm sublayer thickness (not shown) confirm these findings. For smaller sublayer thicknesses, the amorphous/crystalline multilayer structure was observed down to an average sublayer thickness of 15 nm (Fig. 3b-c). However, for average sublayer thicknesses of less than 15 nm, no multilayer structure was observed in TEM (Fig. 3d). For some samples, the sublayer thicknesses in the TEM micrographs deviate significantly from the average sublayer thickness. This can be attributed to local fluctuations of the sublayer thickness as well as to non-perpendicularity of the TEM cross-sections.

For an average sublayer thickness of 290 nm, 14 at.-% P and 3 at.-% P were measured using EDX in the TEM for the amorphous and crystalline sublayers, respectively. For the sample shown in Fig. 3c having a sublayer thickness of 30 nm, measurements in the amorphous layers revealed 11 at.-% P whereas 4 at.-% P were measured for the crystalline parts of the sample.

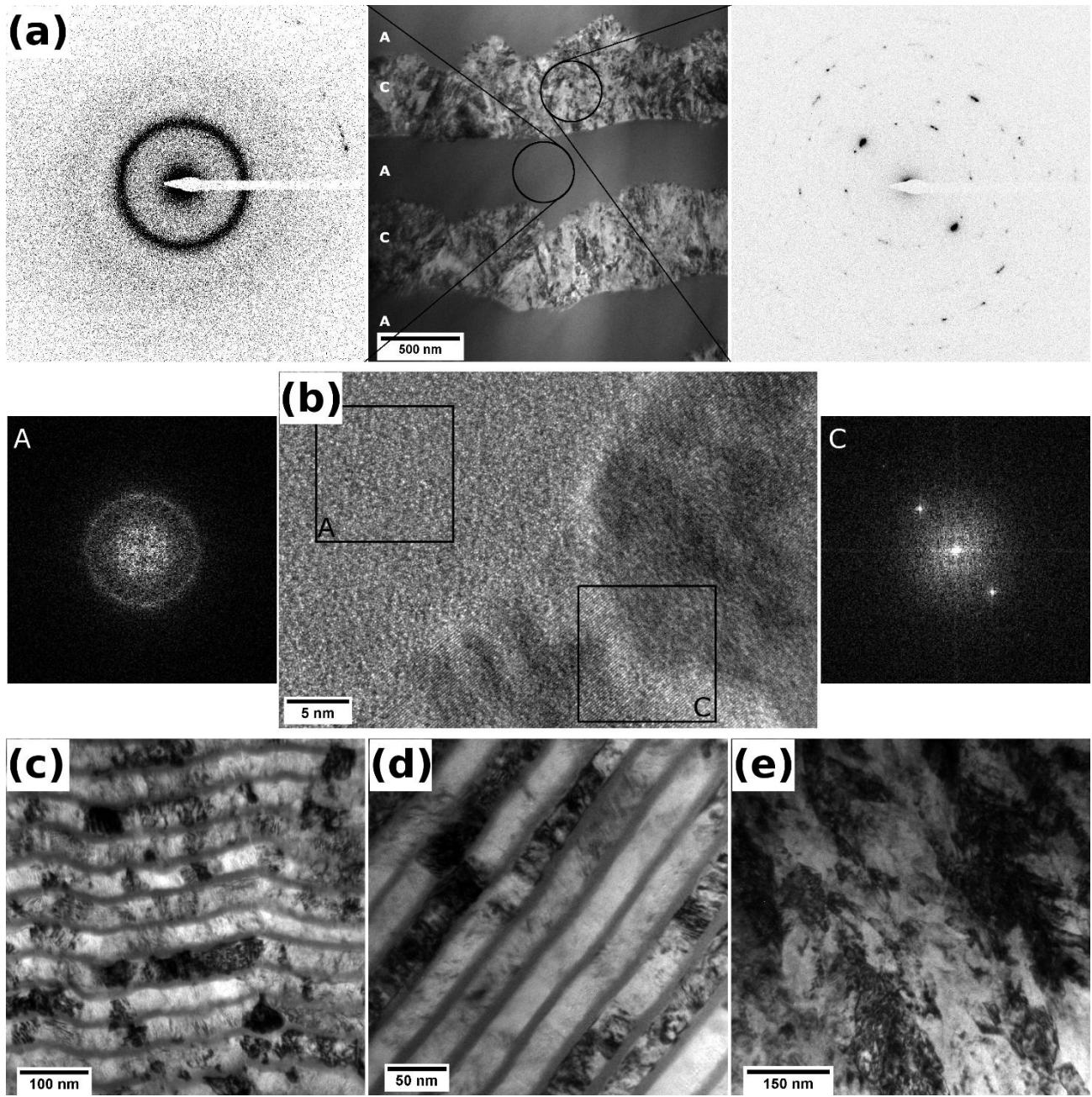

Fig. 3: TEM images of cross-sections: (a) Bright field micrograph and two selected area diffraction patterns of the indicated areas from a sample with 250 nm average sublayer thickness. The amorphous and crystalline sublayers are indicated as A and C, respectively. (b) High-resolution image of an interface in a sample with 290 nm sublayer thickness. The FFT patterns prove the amorphous (left) and crystalline (right) nature of the indicated areas. (c)-(e): Bright field micrographs of samples with (c) 30 nm, (d) 15 nm, and (e) 8 nm average sublayer thickness.

The roughness of the interfaces was evaluated from SEM cross-sections. The interface roughness increases during growth and reaches $R_{rms}$ = 50 nm to $R_{rms}$ = 100 nm at a distance of ten micrometers from the substrate. The final roughness at the surface scatters between $R_{rms}$ = 100 nm and $R_{rms}$ = 250 nm without any obvious dependence on the sublayer thickness.

Residual stress measurements on selected samples using XRD regularly resulted in non-linear $\sin^2\psi$ plots, which may occur due to texture or stress gradient effects [41]. However, more sophisticated measurements are necessary to measure the absolute residual stresses in the crystalline set of sublayers, which is beyond the scope of this work. Nevertheless, all $\sin^2\psi$ plots have a positive slope and $\sin^2\psi$ analysis indicates tensile residual stresses in the order of several hundred megapascals.

3.3. Microhardness and nanoindentation

Fig. 4a-b show the microhardness as a function of the sublayer thickness. The microhardness increases with decreasing sublayer thickness down to an average sublayer thickness of 15 nm, whereas below this value a hardness plateau is obtained. The hardness increase is approximately linear with the inverse square root of the sublayer thickness (Hall-Petch plot, Fig. 4a), but also with the inverse sublayer thickness (Fig. 4b). The former fit has the higher Pearson correlation coefficient, whereas the latter results in the better coefficient of determination (Table 2). Thus, both models describe the measured data with a similar quality.

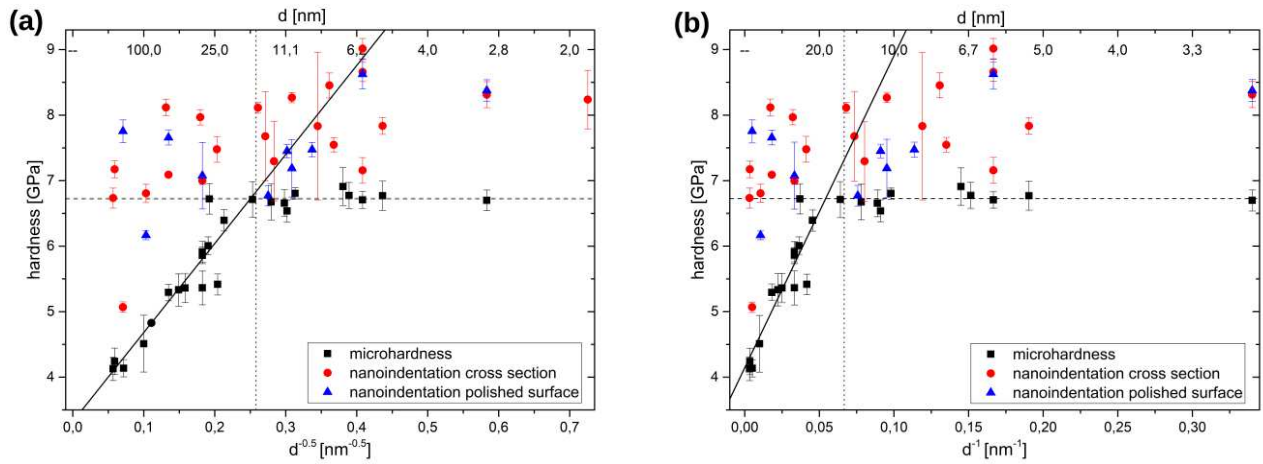

Fig. 4: Microhardness and nanoindentation results plotted (a) in a Hall-Petch plot with respect to the average sublayer thickness and (b) as a function of the inverse average sublayer thickness. The dotted lines show the limit of a sublayer thickness of 15 nm, whereas the dashed lines show the hardness plateau beyond this point.

| Fitting model | Hardness vs. $d^{-0.5}$ | Hardness vs. $d^{-1}$ |
|---|---|---|
| Slope | 13.6 ± 1.4 GPa nm$^{0.5}$ | 47.6 ± 5.8 GPa nm |
| Intercept | 3.3 ± 0.2 GPa | 4.1 ± 0.2 GPa |
| Pearson R | 0.937 | 0.915 |
| R² | 0.878 | 0.837 |

Table 2: Results of the linear fits from Fig. 4a+b and the corresponding Pearson correlation coefficients and coefficients of determination (R²).

The lowest measured hardness value of 4.13 GPa corresponds to an indention depth of 3.0 μm. This is more than the generally recommended 10% of the total layer thickness. However, measurements with reduced load resulted in more scatter due to the smaller indent size, but no significant change in hardness. Thus, no effect of the total layer thickness, i.e. the substrate, on the measured hardness values is expected.

The nanoindentation results scatter significantly more as compared to the microhardness (Fig. 4a-b). However, a similar trend of a hardness increase up to a plateau with decreasing sublayer thickness is observed. Fitting the data in the range of increasing microhardness also results in a positive slope, i.e. an increase in hardness with decreasing sublayer thickness. However, the error of the slope is too large to extract any further information from it. The plateau which is visible in the microhardness data, is almost indistinguishable and only indicated by the data point from the very thinnest layers in the nanoindentation results. Interestingly, no significant deviations between the two investigated loading directions (normal and parallel to the surface normal) occur. All indention curves are smooth without any pop-ins that would be expected if the indentation had initiated delamination along sublayer interfaces (Fig. 5a).

Since the Hall-Petch behavior is often observed in crystalline, but not in amorphous materials, additional hardness measurements were carried out on samples with a fixed nominal crystalline sublayer thickness of 100 nm and a nominal amorphous sublayer thickness varying from 25 to 300 nm. The results show no dependence of the microhardness on the thickness of the amorphous sublayer, but depend more on the scatter of the actual crystalline sublayer thickness (Fig. 5b).

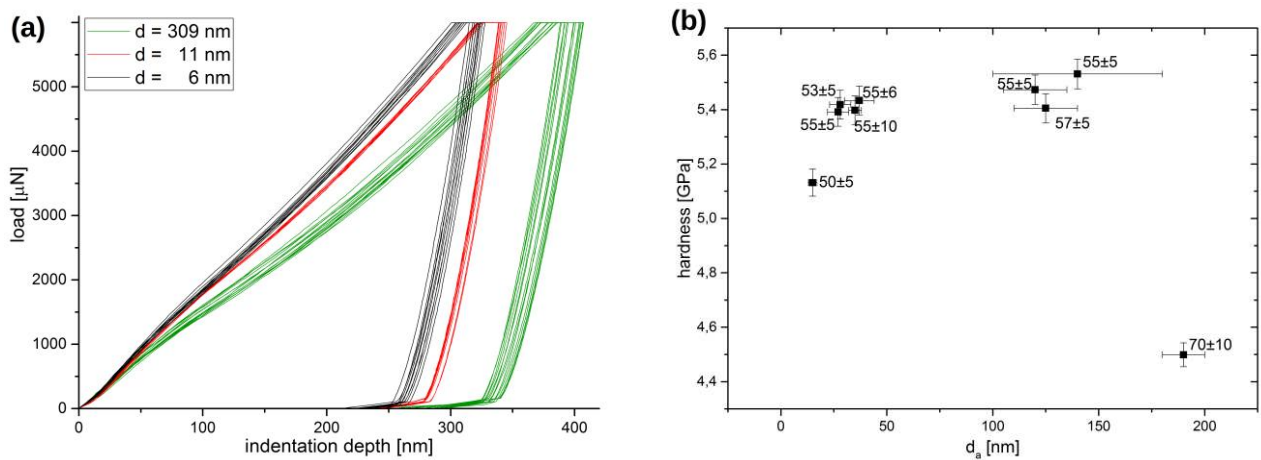

Fig 5: (a) Exemplary nanoindentation curves of samples with 6, 11, and 309 nm sublayer width, d, which show no sudden displacement bursts (pop-ins). (b) Microhardness measurements of samples with a constant nominal crystalline sublayer thickness of 100 nm and varying amorphous sublayer thickness, $d_a$. The numbers indicate the crystalline sublayer thicknesses. All sublayer thicknesses were measured on etched cross-sections in the SEM.

3.4. Microbending tests

Fig. 6a shows the normalized load-displacement curves from the microbending experiments, which were performed on samples with 5 nm, 40 nm, and 310 nm average sublayer thickness. Despite the larger scatter of the data, the highest normalized load, i.e. the highest stresses, are obtained for the 40 nm sample, followed by the 5 nm and finally the 310 nm sample. Despite the highest stresses, also a deviation from a linear load-displacement relationship exists for some of the bending beams from the 40 nm sample. The in-situ SEM observation during the bending experiments of this sample revealed, that a crack is first initiated at the top surface of the bending beam and subsequently grows through the beam upon further loading (Fig. 6b, video S1c). In contrast, the bending beams of the 5 nm and 310 nm samples all break through instantaneously after the crack was observed (videos S1a-b).

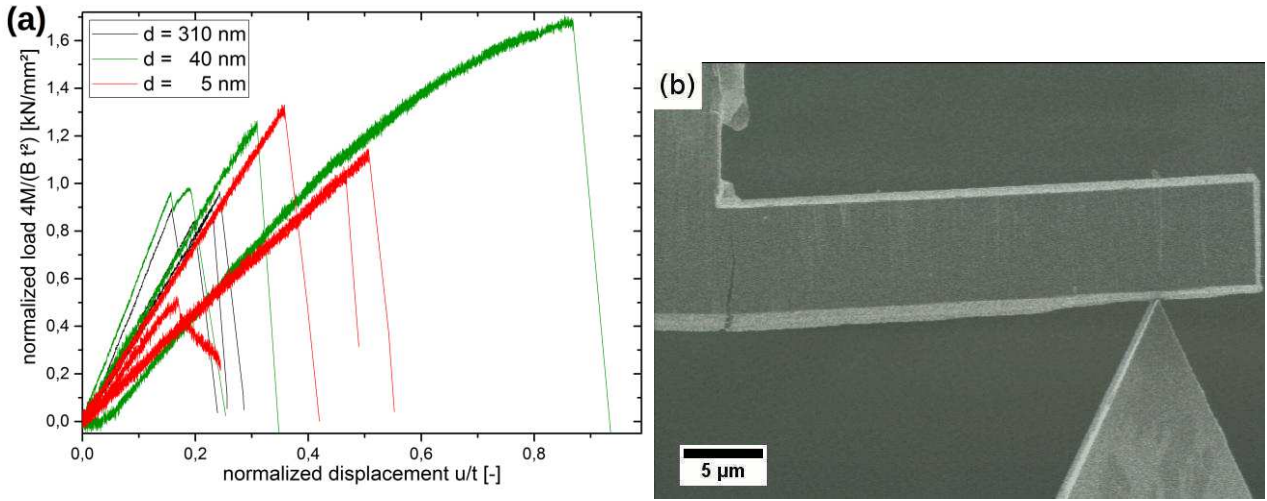

Fig 6: (a) Normalized force displacement diagram of the microbending tests. (b) In-situ SEM micrograph during the microbending experiment of the 40 nm sample with growing crack.

All bending beams failed parallel to the loading direction. The fracture surfaces were investigated with the SEM to gain information about the crack path. For the 310 nm sample, the fracture surface consists of alternating layers of small dimples and relatively flat regions (Fig. 7a). The thickness of these alternating layers corresponds well with the sublayer thickness of the multilayer structure. EDX measurements revealed a higher phosphorus content in the flat regions of the fracture surface, indicating that these regions correspond to the amorphous sublayers. However, absolute phosphorus contents could not be obtained due to the large interaction volume for EDX measurements in the SEM. In the fracture surface of the 40 nm sample, a layered structure was also observed, but only in a part of the fracture surface (Fig. 7c+d). Again, the layer thickness equals approximately the sublayer thickness of the multilayer structure. However, no amorphous and crystalline layers could be identified due to the smaller layer thickness. The fracture surface indicates that the crack was deflected at many layer interfaces, but finally kept its main growth direction parallel to the loading axis leaving no indications of the multilayer structure on the final part of the fracture surface. On the contrary, no layered structure was observed on the fracture surface of the 5 nm sample, which coincides with the loss of the multilayer structure for this sublayer thickness according to the TEM investigations (Fig. 7b).

12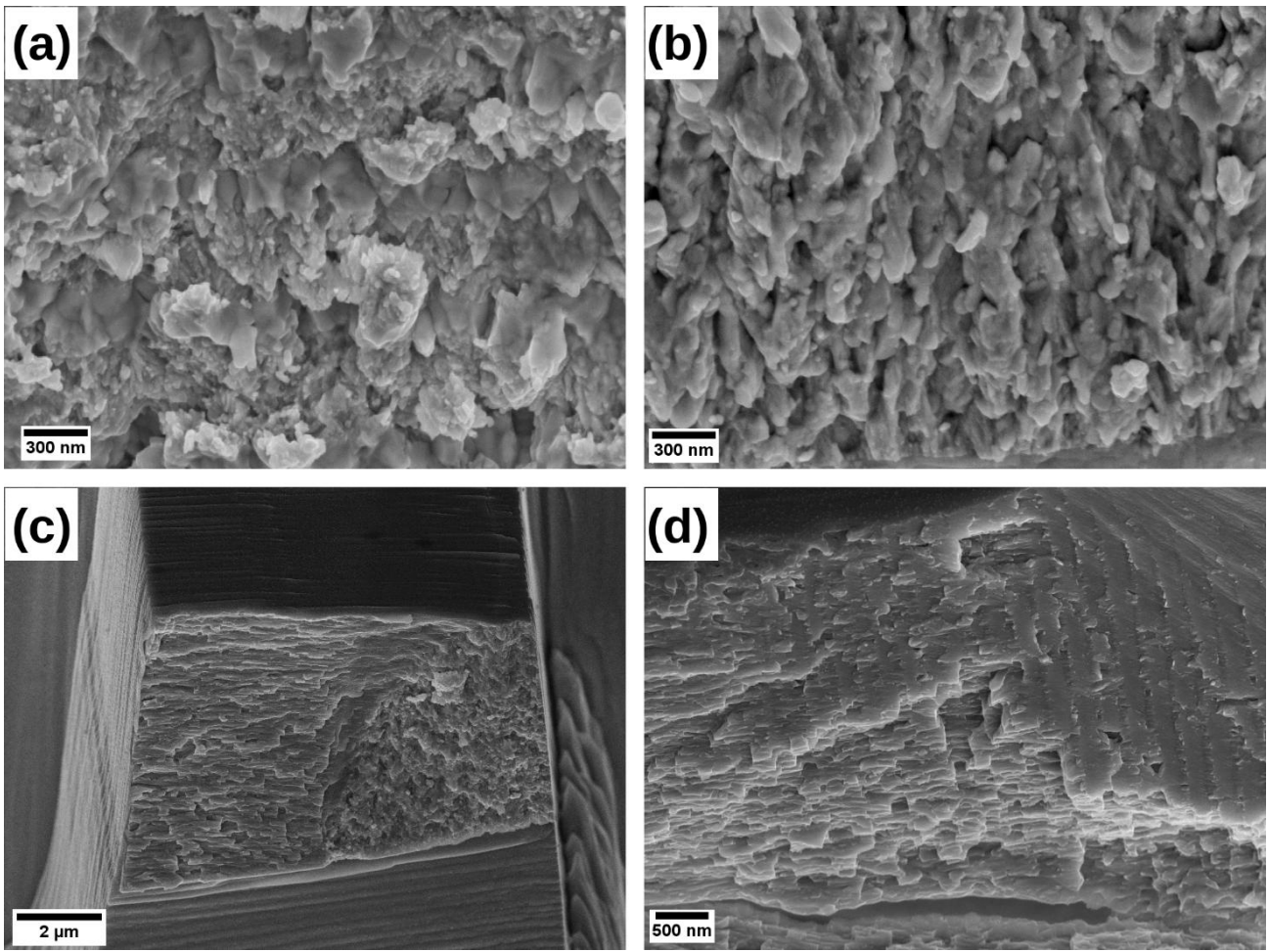

Fig. 7: SEM images of the fracture surfaces of the microbending beams with (a) 310 nm, (b) 5 nm, and (c+d) 40 nm sublayer thickness. The images were taken with the primary electron beam approximately 30° (a+b) or 45° (c+d) to the fracture surface.

4. Discussion

4.1. Hall-Petch behavior for larger layer thicknesses

For the microhardness data of this study, both the Hall-Petch relation and a linear behavior with the inverse sublayer thickness describe the data with similar accuracy. Also in literature, both linear dependence on the inverse square root sublayer thickness (Hall-Petch) or on the inverse sublayer thickness are reported [7-12]. In the following, the results of the Hall-Petch fit are analyzed quantitatively and discussed in detail.

The hardness of the multilayer structure is assumed to combine the hardness of the two phases via a linear rule of mixture. The hardness of the amorphous phase is independent of the layer thickness, as it is observed in amorphous/amorphous multilayers [14]. It may change if its composition changes with varying sublayer thickness. Although the results from the EDX measurements in the TEM indicate such a change, it will not be taken into account in the following analysis. On the one hand, even some signal from the substrate was observed in the EDX measurements indicating that the signal originates not completely from the intended area of interest. This explains the slightly lower phosphorus contents as compared to literature values for amorphous Fe-P alloys as well as the decreasing phosphorus content with decreasing sublayer thickness since more low-phosphorus crystalline material is in close vicinity of the measurement area. On the other hand, slight variations in compositions are not expected to have such a significant effect as in crystalline matter. For these reasons, the hardness of the amorphous structure is set to a constant value of 4.65±0.18 GPa as measured for the amorphous samples produced with direct current (Table 1). On the other hand, the hardness of the crystalline phase is assumed to depend on the sublayer thickness according to the Hall-Petch relation. With these assumptions, an intercept of 1.99±0.40 GPa and a slope of 859.2±88.8 MPa $\mu m^{0.5}$ are obtained for the Hall-Petch equation of the crystalline phase. Using a conversion factor of three [42], this equals $\sigma_0$ = 660±130 MPa and $k_{HP}$ = 286±30 MPa $\mu m^{0.5}$ for the yield stress. The intercept is significantly lower than the hardness of the crystalline single phase as obtained from the direct current samples (Table 1). However, possible deviations of the structure in a sublayer of a multilayer structure and in a direct current sample, which are prepared with the same deposition parameters, have to be considered. First, during the deposition of a crystalline sublayer on a multilayer structure, that has already a certain total thickness, less phosphorus is expected to be incorporated as compared to the deposition with the same parameters on a direct current sample of the same thickness, since the concentration of hypophosphite in the electrolyte close to the substrate is lower due to the larger previous consumption. This change in phosphorus content directly affects the mechanical properties due to solid solution hardening. Secondly, the high current density and the enrichment in phosphorus in the direct current samples result in a grain size that is significantly too small to take these samples as materials with quasi-infinite grain size as necessary for the intercept of the Hall-Petch plot. In most interstitial-free steels, the yield stress is much lower than the results of this analysis, but they contain only a small amount of phosphorus [42]. In solution treated Fe-P alloys a similar hardness to the hardness intercept in the present study was obtained with about 1 to 2 at.-% P [44], which is close to the compositions measured with EDX in the TEM. These measurements also indicate that there is no significant change of composition with varying sublayer thickness since the difference from 3 at.-% to 4 at.-% going from 290 nm to 30 nm sublayer thickness is close to the measurement inaccuracy of EDX.

The obtained Hall-Petch constant of 286±30 MPa $\mu m^{0.5}$ for the yield stress seems to be too large for Fe-P alloys since the Hall-Petch constant for interstitial-free steels is reported to be between 130 and 180 MPa $\mu m^{0.5}$ [45] and is slightly lowered by the presence of phosphorus [46]. The presence of interstitial carbon would increase the Hall-Petch constant significantly, even for very low concentrations [46]. However, sodium dodecyl sulfate, which is the only component of the electrolyte that contains carbon, is present just in very low concentrations and is not known to result in co-deposition of carbon. Thus, the presence of carbon in the deposits can be excluded. The uncertainty of the conversion factor between hardness and yield strength might also be another reason for the discrepancy, since the conversion factor of three is just an approximation [42] and a large range of conversion factors is reported in literature [47, 48]. However, the main reason for the too large Hall-Petch constant might be the assumption of equal thickness of amorphous and crystalline sublayers for the determination of the sublayer thicknesses. The TEM results indicate that the assumption is approximately correct for large sublayer thicknesses, but the amorphous sublayers are much thinner than the crystalline ones for smaller thicknesses (see Fig. 3). Thus, the phase fraction of the crystalline phase increases with decreasing sublayer width, resulting in an additional increase of strength beside the Hall-Petch effect as soon as the crystalline phase is harder than the amorphous phase. This effect was not included in the analysis above, which results in a too large Hall-Petch constant. However, a quantitative analysis of this effect would require the accurate measurement of the crystalline sublayer thicknesses for all samples which is impossible due to local fluctuations, calling for an unmanageable number of TEM samples, and the effect of etching parameters on the thickness of the individual layers for SEM investigations. Furthermore, for the largest sublayer thicknesses, grains smaller than the sublayer thickness are visible in the TEM micrographs (Fig. 3a); in this case the hardness could be additionally influenced by the grain size. However, there are at least some grains going through the complete crystalline sublayer for all sublayer thicknesses. Thus, the maximum grain size in growth direction of the films always equals the sublayer thickness. For this reason, we do not expect a significant influence of grain size in our samples. However, a detailed grain size analysis was not performed and is out of the scope of this study. To separate the effects of grain size and sublayer thickness, one had to produce samples with constant sublayer thickness but different grain sizes. This may be achieved by modifying the deposition parameters. However, this might induce a simultaneous change in composition resulting again in two different hardening mechanisms which are difficult to separate.

4.2. Breakdown of the Hall-Petch relation and loss of multilayer structure for thinner layers

Transitions from Hall-Petch behavior to a hardness (or yield strength) plateau at a grain size or sublayer thickness in the nanometer regime are frequently reported. They are usually attributed to a change in deformation mechanism due to a threshold size below which no dislocation pile-up at the grain boundaries or multilayer interfaces can form [9, 10]. This cannot be excluded for the transition observed in this study. However, from the TEM images in Fig. 3 another reason for the restricted range of validity of the Hall-Petch relation in the present case can be suggested: the loss of the amorphous/crystalline multilayer structure. The absence of continuous amorphous sublayers for average sublayer thicknesses smaller than 15 nm allows crystallites to grow to a size much larger than the intended sublayer thickness. The presence of multilayer structures in some regions of the samples in the SEM investigations indicate that there is still a periodic modulation (probably of phosphorus content) which results in a periodic change of the attack by the etching solution. Two main reasons may result in the loss of the amorphous sublayers for short periodicity lengths: the interface roughness and the decreasing amorphous-to-crystalline phase ratio with decreasing sublayer thickness. The roughness causes an inhomogeneous current density distribution during the deposition process. Once this inhomogeneity becomes too large, the current density during the deposition of the amorphous sublayers becomes sufficiently large to deposit not an amorphous, but a crystalline alloy at the regions, where the current density distribution has its maxima. The opposite effect during the deposition of crystalline sublayers at the minima of the current density distribution is less likely, since the minima coincide with the lowest points in the surface profile of the deposit, which are surrounded by an electrolyte depleted in hypophosphite due to geometrically impeded mass transport or diffusion, respectively. The decreased ratio of amorphous-to-crystalline sublayer thickness enhances the proposed mechanism for the breaking of the multilayer structure, since the thinner amorphous layers facilitate the penetration by crystalline phase.

4.3. Microbending tests and nanoindentation

The results in both microbending and nanoindentation scatter more than the microhardness measurements. This can be attributed to the smaller volume which is tested in these small-scale techniques. When local defects such as small pores or inclusions (e.g. hydroxides or phosphides) on nanometer scale are present in the tested volume of such tests, they result in measurable reduction in strength or hardness, respectively. On the other hand, a larger volume is probed in microhardness testing and the obtained values are averaged over a larger volume. Additionally, roughness and local deviations in sublayer thickness also attribute to the larger scatter in the testing methods with a smaller testing volume.

Despite the scattering, the microbending tests indicate some ductility for the intermediate nominal sublayer thicknesses, for which the sublayer thickness is already in the nanometer regime, but the amorphous/crystalline multilayer structure is still present. This can be attributed to the multiple crack deflection at the large number of interfaces, as visible on the fracture surfaces.

However, the maximum stresses obtained in microbending are lower than expected from the hardness measurements. Besides local defects, as discussed above, the residual stresses in the deposits are another reason for this discrepancy. Tensile residual stresses for the crystalline phase, that were obtained from X-ray stress analysis, are a typical feature of iron electrodeposits [49] and have also been found in purely amorphous Fe-P deposits [29]. They promote crack initiation and crack growth in the tension part of the bending beam. On the other hand, their influence in the complex stress field during indentation is more complicated, but is expected to be less significant since the stresses are mainly compressive and the strains are quite large in the vicinity of the indent. The tensile residual stresses can also cause the observed delamination in multilayers with small sublayer widths after etching (Fig. 2d), since the etched region acts as a notch, which results in crack growth or delamination under the tensile residual stresses if the stress concentration is high, i.e. if the notch is sharp enough. Thus, further improvement of the mechanical properties may be obtained by a reduction of residual stresses, which can be reached by changes in the deposition process (e.g. addition of suitable additives or pulse plating).

## 5. Summary

It was shown that amorphous/crystalline multilayer structures of Fe-P alloys can be prepared using the single bath electrodeposition technique. However, the multilayer structure breaks down below a sublayer thickness of 15 nm which is attributed to the role of roughness and amorphous-to-crystalline layer thickness ratio during the deposition process. A significant increase in strength according to the Hall-Petch relation was observed in the multilayer region, whereas a hardness plateau was observed when the multilayer structure was broken. Microbending tests showed that high strength and crack deflection at the interfaces can be obtained. Further improvement of the mechanical properties may be achievable in future by appropriate changes in the deposition process in order to decrease the minimum sublayer thickness for intact amorphous/crystalline multilayer structures as well as the number of local defects and the amount of residual stresses.

## Acknowledgments

The authors would like to thank Karoline Kormout (Erich Schmid Institute, Austrian Academy of Sciences) for help with TEM experiments and for performing the EDX and high-resolution measurements. We gratefully acknowledge the financial support by The European Research Council under ERC Grant Agreement No. 3401 85 USMS and by the Austrian Science Fund (FWF): J3468-N20.

## Supplementary Material:

S1: videos of in-situ SEM observation of microbending experiments at three bending beams with (a) 310 nm, (b) 5 nm and (c) 40 nm average sublayer thickness.